# Normalizing and Classifying Shape Indexes of Cities by Ideas from Fractals

Yanguang Chen

(Department of Geography, College of Urban and Environmental Sciences, Peking University, Beijing 100871, P.R. China. E-mail: chenyg@pku.edu.cn)

**Abstract**: A standard scientific study comprises two processes: one is to describe a thing, and the other is to understand how the thing works. In order to understand the principle of urban growth, a number of shape indexes are proposed to describe the size and shape of cities. However, the comparability of a shape index is often influenced by the resolution of remote sensing images or digital maps because the calculated values depend on spatial measurement scales. This paper is devoted to exploring the scaling in classical shape indexes with mathematical methods. Two typical regular fractals, Koch's island and Vicsek's figure, are employed to illustrate the property of scale dependence of many shape indexes. A set of formula are derived from the geometric measure relation to associate fractal dimension for boundary lines with shape indexes. Based on the ideas of fractals, two main problems are solved in this work. First, several different correlated shape indexes such as circularity ratios and compactness ratios are normalized and unified by substituting Feret's diameter for the major axis of urban figure. Second, three levels of scaling in shape indexes are revealed, and shape indexes are classified into three groups. The significance of this study lies in two aspects. One is helpful for realizing the deep structure of urban form, and the other is useful for clarifying the spheres of application of different types of shape indexes.

## 1. Introduction

Effective description lays the foundation for an academic study, as is known to scientists. Scientific research should proceed first by describing how a system works and then by understanding why (Gordon, 2005). The historian of science, Henry (2002, page 14), once made comments on scientific method as below: "the two main elements of this scientific method are the use of mathematics and measurement to give precise determinations of how the world and its parts work, and the use of observation, experience, and where necessary, artificially constructed experiments, to gain understanding of nature." In short, a study process comprises two steps: one is to describe a thing using words, numbers, mathematics, and measurement, the other is to understand how the thing works by means of observation, experience, experiments, simulation, computation, and so on. In social science, the data based on questionnaire survey is always employed to find macro regularity (for description), while the interview method is adopted to reveal the micro mechanism (for understanding). In urban studies, it is necessary to describe the size and shape in the growth of human settlements (Batty, 2008; Batty and Longley, 1987; Chen, 2011; Dutton, 1973; Longley *et al*, 1991; Medda *et al*, 1998; Wentz, 2000). Modeling the relationships between size and shape of cities and towns is a basic work for urban geography. An important approach to describing urban shape is to construct some kind of index using the basic measures such as length, area, and point number. In fact, a number of elementary indexes are successively presented to characterize the shape of geographic area (Haggett *et al*, 1977). The great majority of these indexes can be employed to describe urban shape. A shape index is actually of spatial measurement. On the one hand, measurement makes the numerical link between mathematical modeling and empirical research (Taylor, 1983). On the other, an index provides a simple figure in which much numerical properties of a thing are concentrated (Chen, 2016; Wheelan, 2013). Form represents the pattern of force (Thompson, 1966). Urban form is always reflected by shape, size, scale, and density distribution. By comparing the shape indexes of different spatial and temporal elements, we can obtain significant geographical information of urban form and systems of cities.

Unfortunately, the shape indexes are often ineffective owing to the uncertainty of spatial measurements such as perimeter and area. The index values of different times or places are always not comparable with one another. For the same city at the same time, the calculation results based on the remote sensing images of different resolution ratios are different. In order to solve the problem of uncertainty of measurement, more shape indexes were invented. However, before appearance of fractal theory, the fundamental problem cannot be solved. The reason lies in the scale invariance of urban form. Geographical phenomena such as cities can be divided into two types:

one is phenomena with characteristic scale and can be described with traditional mathematic methods and measurement, the other is those without characteristic scale and cannot be effectively described by traditional methods (Chen, 2015a). No characteristic scale indicates some kind of scaling. Scaling is to reduce or enlarge a thing according to a fixed ratio (Williams, 1997). If the scaling process change the size but does not change the shape and structure of a thing, we meet with a scaling law. In other words, scaling law implies invariance of properties, patterns, or relation in a process of transformation. For cities, scaling law suggests the scale dependence of spatial measurement. One of the powerful tools for scaling analysis is fractal geometry, which can be employed to improve urban measurement (Batty and Longley, 1994; Frankhauser, 2008; Thomas *et al*, 2008). The uncertainty of geographical measurements is always associated with fractal properties of geographical systems (Batty and Longley, 1994; Chen, 2008; Chen, 2011; Frankhauser, 1994; Thomas *et al*, 2010). Fractals suggests the optimum structure of both natural and human systems. A fractal object can occupy its space in the best way. Using the ideas from fractals to plan cities and systems of cities, we will be able to make the most of the geographical environment and natural resources. The precondition of fractal planning is to develop the theory of fractal cities, and the precondition of developing fractal theory of cities is to find effective methods of spatial description and understanding. This paper is devoted to exploring the scaling in the classical shape indexes of cities, which will suggests new way of describing urban shape and form. In Section 2, a simple fractal line, Koch's island, is employed to illustrate the scaling of shape indexes. Then, the common shape indexes are normalized and classified by the concepts from Feret's diameter and scaling. In Section 3, simple case analyses based on two typical regular fractals are made to show the three levels of scaling properties of shape indexes. In Section 4, several questions are discussed. Finally, the paper is concluded with a summary of this study.

## 2. Models

### 2.1 Koch island as a metaphor of urban shape

Simple examples and clear cases are helpful for understanding complex systems and sophisticated problems. Koch curve is a well-known simple fractal line, and a boundary of Koch's island comprises three Koch curves that form a closed region. The area confined by the three Koch curves is termed Koch island, which is sometimes confused with Koch snowflake. There is an analogy between urban boundary and Koch curve (Batty and Longley, 1994). Thus, Koch's island can be employed to model urban area. Taking an equilateral triangle as an initiator, we can generate a Koch's island by an iterative process (Figure 1). Apparently, the boundary is a fractal line, but the area within the boundary is a Euclidean surface.

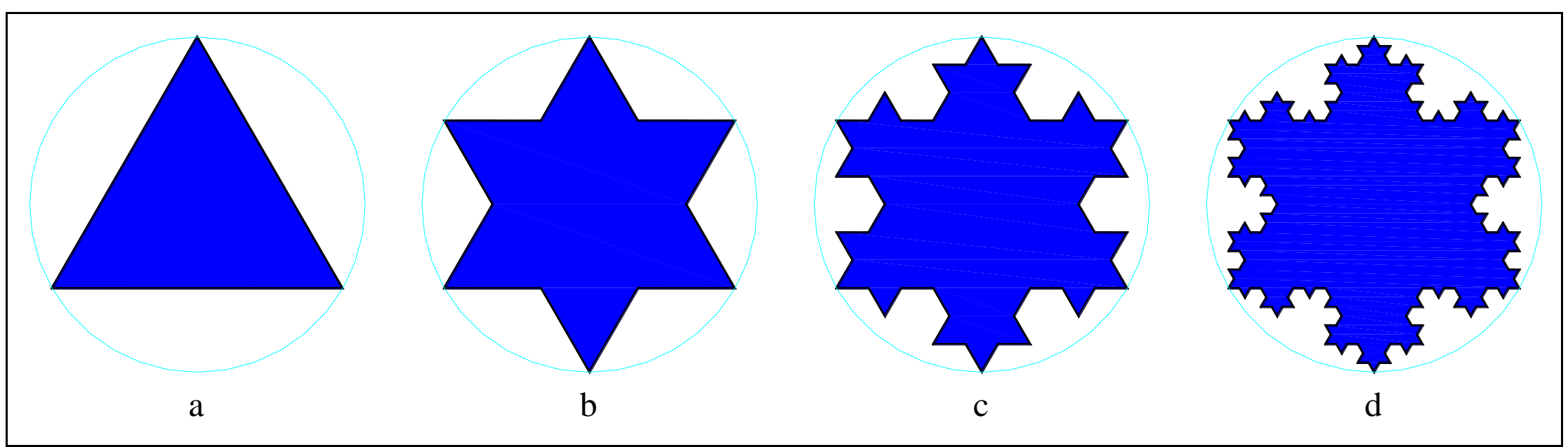

**Figure 1 A schematic diagram of the Koch island generation (The first four steps)**

Fractal lines suggest the simplest way of understanding fractal measurements. A number of early studies on fractal cities started from geographical fractal lines (e.g. Batty and Longley, 1988; De Keersmaecker *et al*, 2003; Longley and Batty, 1989a; Longley and Batty, 1989b). If we use a ruler, the length of which is $\varepsilon$, to measure the length of a Koch curve, and the times of measurements will be $N(\varepsilon)$, then the relation between the linear size $\varepsilon$ and the times $N(\varepsilon)$ will form a power function as below

$$N(\varepsilon) = N_1 \varepsilon^{-D}, \tag{1}$$

where $N_1$ and $D$ are two parameters: proportionality coefficient and fractal dimension (Mandelbrot, 1982). In theory, $N_1$=1, and $D$=ln(4)/ln(3)≈1.2619. The perimeter of Koch's island can be estimated by the following formula

$$P(\varepsilon) = 3N(\varepsilon)\varepsilon = 3N_1 \varepsilon^{1-D}. \tag{2}$$

This implies that if the linear size approaches to infinitesimal, the perimeter will approach infinity. On the other hand, if the area of the initial triangle is $A_0$=1, then the terminal area of the Koch island will approach a constant, that is

$$A = A_0[1 + \frac{1}{3}\lim_{m\to\infty}\sum_{i=1}^{m}(\frac{4}{9})^{i-1}] = 1 + \frac{1}{3-4/3} = \frac{8}{5}. \tag{3}$$

Further, if we define a shape index of Koch's island, the index can be proportional to the ratio of the island area to the square of the island's circumference, i.e., $A/P^2$. Since $N_1$=1, we have

$$ratio = \frac{A(\varepsilon)}{P(\varepsilon)^2} = \frac{8}{45\varepsilon^{2(1-D)}} . \tag{4}$$

As for $A^{1/2}/P$, the ratio can be expresses as

$$ratio = \frac{\sqrt{A(\varepsilon)}}{P(\varepsilon)} = \frac{2\sqrt{2/5}}{3\varepsilon^{1-D}} . \tag{5}$$

The area is a finite value, while the perimeter is infinite. The ratios will become an infinitely small quantity. The shape index based on the ratios will depend on the scale of measurement. However, if the shape index is based on the ratio of the area to the length of major axis, the value will be determinate because that the area of Koch island is certain.

## 2.2 Integration of shape indexes by Feret's diameter

A number of shape indexes have been proposed to characterize the surface configuration of geographical area. The main indexes include circularity ratios (CiR), Compactness ratios (CoR), Ellipticity index (EI), Elongation ratio (ER), Form ratio (FR), Radial shape index (RSI), and so on (Chen, 2011; Haggett *et al*, 1977; Lin, 1998; Taylor, 1983). For a figure, these indexes are based on the following basic measures: figure area ($A$), longest (major) axis ($L$), secondary (minor) axis ($L'$), perimeter ($P$), area of minimum circumscribed circle ($A'$), and the distance of the $i$th vertex from centroid of the figure ($d_i$) (Table 1). If we have a set of normalized concepts of urban size and shapes, these shape indexes can be principally integrated into a coherent framework. In fact, four basic and significant definitions of cities have been proposed by Longley *et al* (1991) and Batty and Longley (1994) (Figure 2). Using these definitions, we can normalize the shape indexes and find the mathematical links between them.

**Table 1 The classical elementary indices for measuring the shape of geographic area**

| Shape index (SI) | Formula | Author/presenter |
| --- | --- | --- |
| **Circularity ratio (I), CiR** | $4A/P^2$ | **Miller/1953** |
| **Circularity ratio (II), CiR** | $2\sqrt{\pi A}/P$ | **Richardson/1961** |
| **Compactness ratio (I), CoR** | $4A/(\pi L^2)$ | **Gibbs/1961** |
| **Compactness ratio (II), CoR** | $A/A'$ | **Cole/1964** |

| Form ratio, FR | $A/L^2$ | Horton/1932 |
| --- | --- | --- |
| Ellipticity index, EI | $L/\{2A/[\pi(L/2)]\}$ | Stoddart/1965 |
| Radial shape index (II), RSI | $A/\sqrt{2\pi\int d_i^2\mathrm{d}x\mathrm{d}y}$ | Blair and Bliss/1967 |
| Elongation ratio, ER | $L/L'$ | Werrity/1969 |
| Radial shape index (I), RSI | $\sum_{i=1}^{n}\left\|(100d_i/\sum_{i=1}^{n}d_i)-(100/n)\right\|$ | Boyce and Clark/1964 |

**Source**: Haggett *et al*, 1977, Page 309 (a secondary source). **Note**: $A$=area, $L$=longest (major) axis—the longest line between two points on the shape's perimeter, $L'$=secondary (minor) axis, $P$=perimeter, $A'$=area of smallest circle to enclose figure, $d_i$=distance of $i$th area from centroid of the figure.

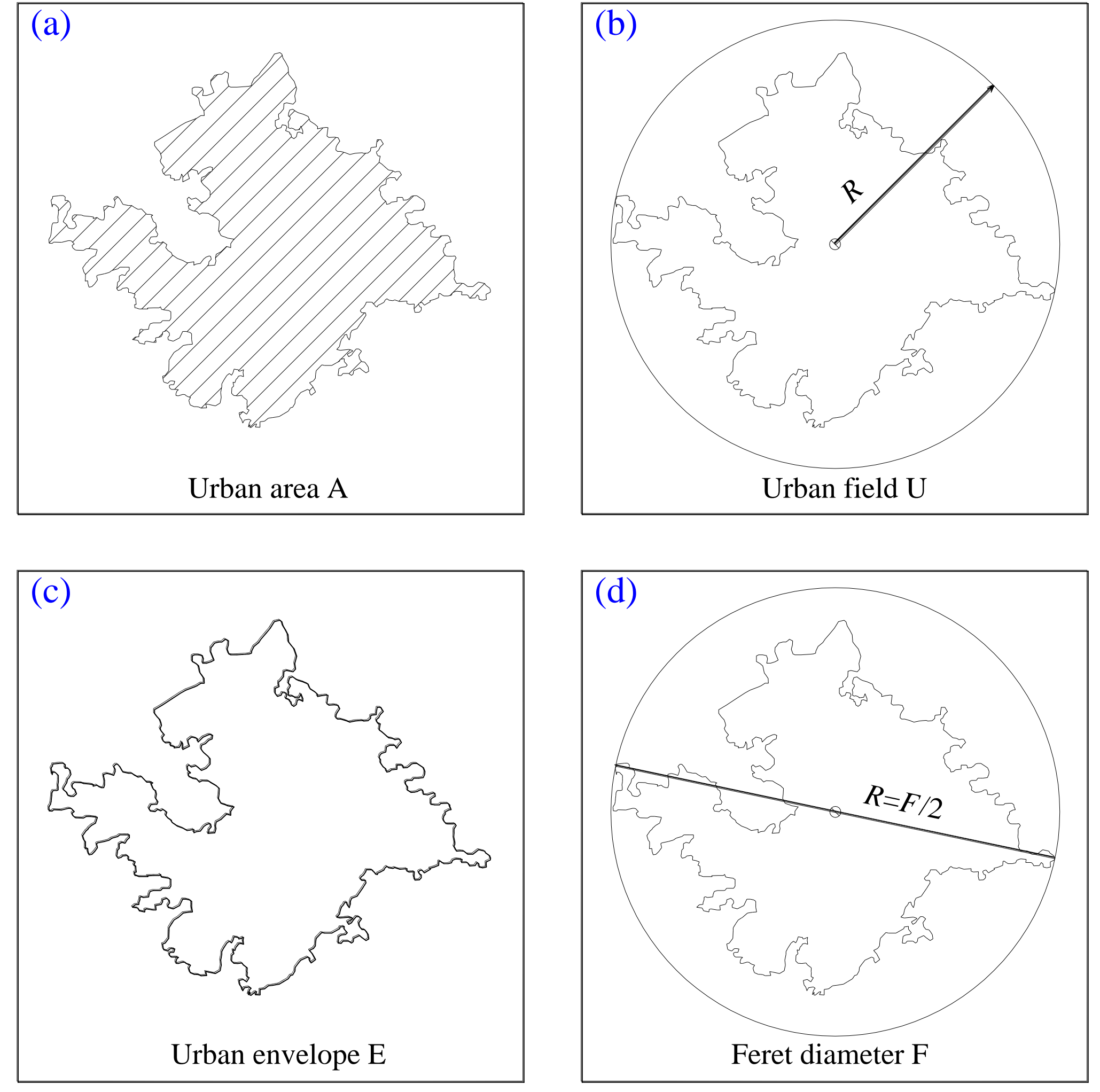

**Figure 2 Definitions of urban area, field, envelope and radius (by** Batty and Longley (1994) **and** Longley *et al* (1991)**)**

The four normalized urban concepts are defined as follows (Batty and Longley, 1994; Longley *et al*, 1991). Drawing the outline of a city, we will have an urban boundary. The boundary curve is termed *urban envelop*, and the area within the urban boundary is termed *urban area*. If we can find the minimum circumscribed circle to enclose the figure of urban envelop, the circular area forms an *urban field*, and the diameter is termed *Feret diameter* (Batty and Longley, 1994; Longley *et al*, 1991; Kaye, 1989). The four concepts have inherent logic and numerical links. The corresponding relationships between the four concepts and the measures used in the traditional shape indexes are tabulated as below (Table 2). The difference lies in the Feret's diameter and the major (longest) axis. For the regular fractal figure, Feret's diameter is just the length of the major axis. For the random figure, however, Feret's diameter is not necessarily equal to the length of the major axis. Generally speaking, the former is close to the latter. Substituting Feret's diameter $F$ for the major axis $L$, we can redefine the minor axis as below. Suppose that there is a line through the center of a circle perpendicular to the Feret diameter. The minor axis is the line segment intercepted by the urban envelope. Based on the new definitions of major and minor axes, the connections between different formulae of shape indexes can be brought to light (Chen, 2011). For example, the first compactness ratio (CoR1) formula, $4A/(\pi L^2)$, is equivalent to the second compactness ratio (CoR2) formula, $A/A'$. According to the definition of basic measurements, we have

$$A' = \pi(\frac{F}{2})^2 = \frac{1}{4}\pi L^2 . \tag{6}$$

So

$$\frac{A}{A'} = A / \frac{1}{4}\pi L^2 = \frac{4A}{\pi L^2} , \tag{7}$$

which suggests that the compactness ratio (CoR) is proportional to the form ratio (FR). The ellipticity index (EI) is just the reciprocal of the compactness ratio (CoR), that is

$$\frac{L}{2A/(\pi L/2)} = \frac{\pi L^2}{4A} = \frac{A'}{A} . \tag{8}$$

These equivalence relationships can be easily testified by next numerical analysis of regular fractals.

**Table 2 The corresponding relationships between the definitions of urban area, field, envelope and radius and the measures in shape indexes**

| Measure | The definition in fractal cities | The measures in shape indexes |
|---|---|---|
| **Area** | Urban area--The area within an urban envelope ($A$) | Urban area--The area within an urban boundary ($A$) |
| **Circumcircle** | Urban field--Minimum circumscribed circle of urban envelope ($U$) | Smallest circumcircle--The area of the smallest circle to enclose figure ($A'$) |
| **Perimeter** | Urban envelope -- Urban boundary ($E$) | Perimeter--The length of urban boundary ($P$) |
| **Diameter/ Longest axis** | Feret's diameter--The diameter of urban field ($F$) | Major axis--The longest line between two points on the shape's perimeter ($L$) |

## 2.3 Classification of shape indexes by scaling

If we don't consider the internal structure and spatial pattern of land use of a city, Koch's island is indeed a good metaphor of urban shape. If we introduce chance factors into the Koch's island, we will generate random Koch curves, and the random Koch's island looks very like a real urban envelope shown on a digital map. In this case, urban boundaries can be treated as fractal lines (Batty and Longley, 1987; Batty and Longley, 1988; De Keersmaecker *et al*, 2003; Longley and Batty, 1989a; Longley and Batty, 1989b; Song *et al*, 2012; Wang *et al*, 2005). Based on the concepts from prefractals and scaling range, we can derive the relation between partial shape indexes and boundary dimension (Chen, 2011). However, if we consider the internal patterns of urban land use, the urban form is fractal, and both the boundary and land use area depend on measurement scale (Benguigui *et al*, 2006; Chen, 2013; Cheng, 1995; Imre, 2006; Imre and Bogaert, 2004). The box dimension of fractal form indicative of urban land use patterns can be defined by (Benguigui *et al*, 2000; Chen and Wang, 2013; Feng and Chen, 2010; Shen, 2002)

$$N(\varepsilon) = N_1 \varepsilon^{-D_f} , \tag{9}$$

where $N(\varepsilon)$ is the number of nonempty boxes of line scale $\varepsilon$, $N_1$ denotes the proportionality coefficient, and $D_f$ refers to the form dimension. Thus the area of the fractal object can be given by

$$A(\varepsilon) = N(\varepsilon)\varepsilon^2 = A_1 \varepsilon^{2-D_f} , \tag{10}$$

in which $A_1$=$N_1$ denotes the proportionality coefficient. For general fractal boundaries, equation (2)

can be rewritten as below

$$P(\varepsilon)=P_1\varepsilon^{1-D_b}, \tag{11}$$

where $P_1$ denotes the proportionality coefficient, and $D_b$ refers to the boundary dimension. For the Koch island, we have $P_1=3N_1=3$. Combining equation (10) and equation (11) yields an inverse allometric relation as follows (Chen, 2013)

$$A(\varepsilon)=\mu P(\varepsilon)^{(2-D_f)/(1-D_b)}, \tag{12}$$

where the proportionality coefficient $\mu=A_1P_1$^[(2-$D_f$)/($D_b$-1)].

Where there is fractal dimension, there is a scaling process; where there is scaling process, there is uncertainty. According to equation (10), urban area is uncertain; and according to equation (11), urban perimeter is uncertain. The scaling of urban form can be examined from three levels. The first level is the scaling of urban perimeter. Urban boundary can be treated as a fractal line (Batty and Longley, 1994). If we don't consider the internal structure and texture of urban land use, the urban area can be treated as Euclidean area. Based on fractal lines and Euclidean area, the fractal measure relation is as follows (Batty and Longley, 1988; Feder, 1988; Mandelbrot, 1982; Takayasu, 1990)

$$A(r)^{1/2}\propto P(r)^{1/D_b}, \tag{13}$$

where $r$ represents rank or linear size of boxes. The dimension of area is $d$=2, and the boundary dimension of perimeter is $D_b$. Based on equation (13), a shape index indicative of edge roughness was proposed by Wentz (2000). The second level is the scaling of both urban perimeter and area. If we take the intraurban structure of urban patterns into account, urban form is involved with "fractal area" (Figure 3). The Lebesgue measure of a real thin fractal is zero, and this indicates that fractal area approaches zero. However, urban form is prefractal rather than a real fractal. A prefractal can be treated as a fractal within certain scaling range. Based on prefractal lines and area, the fractal measure relation is as below (Benguigui *et al*, 2006; Chen, 2013; Cheng, 1995; Imre and Bogaert, 2004)

$$A(r)^{1/D_f}\propto P(r)^{1/D_b}, \tag{14}$$

in which the fractal dimension of area is $D_f$<2, and the boundary dimension of perimeter is $D_b$. Equation (13) is based on equation (11), while equation (14) is based on equations (10) and (11). The fractal measure relations are in fact allometric scaling relations. The third level is involved with variable urban boundary (Chen *et al*, 2015). The first two levels are based on fixed urban boundary, but it is hard to find clear and objective boundary curve for real cities. Recent 10 years, several new methods are invented to identifying urban boundaries (Jiang and Jia, 2011; Rozenfeld *et al*, 2008; Tannier *et al*, 2011). However, all these methods depend on subjective parameter values, which lead to variable urban boundaries (Chen *et al*, 2015; Tan and Chen, 2015). The problem does not rest with the deficiency of these methods, and the essence of variable boundaries lies in scaling behind urban form. Owing to variability of urban boundaries, the major axis, the minor axis, and the distance of *i*th vertex from centroid of the figure, are not certain. As a result, all the shape indexes are not determinate.

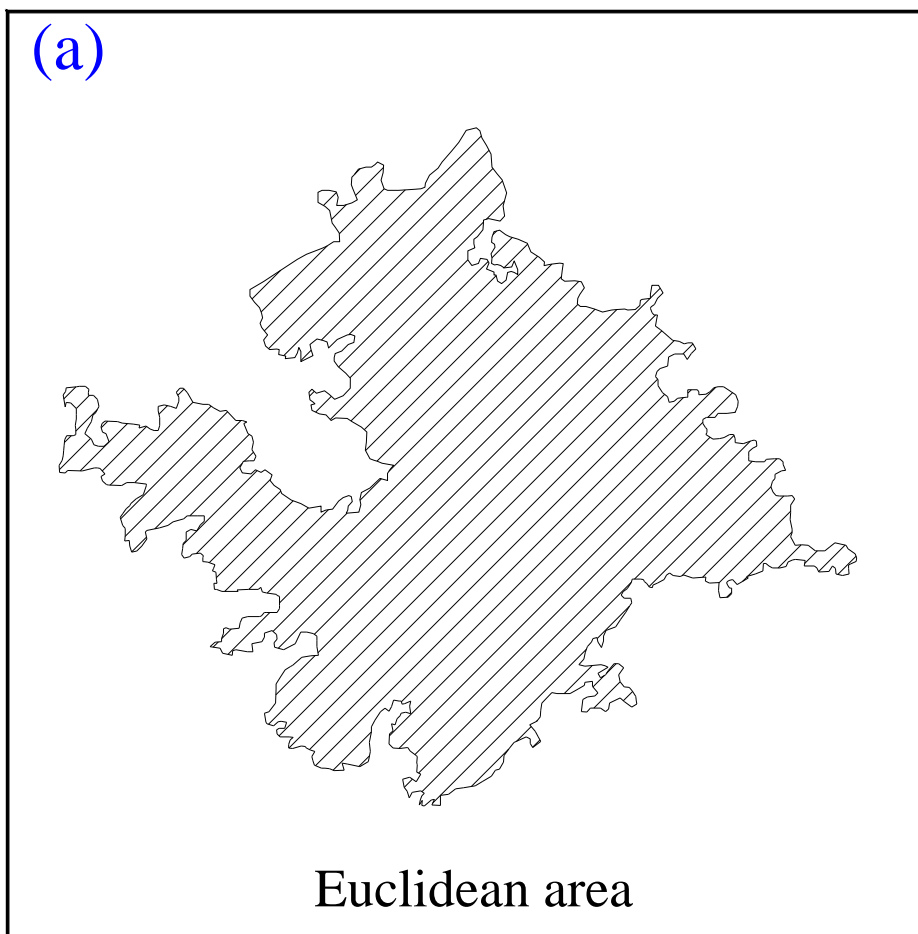

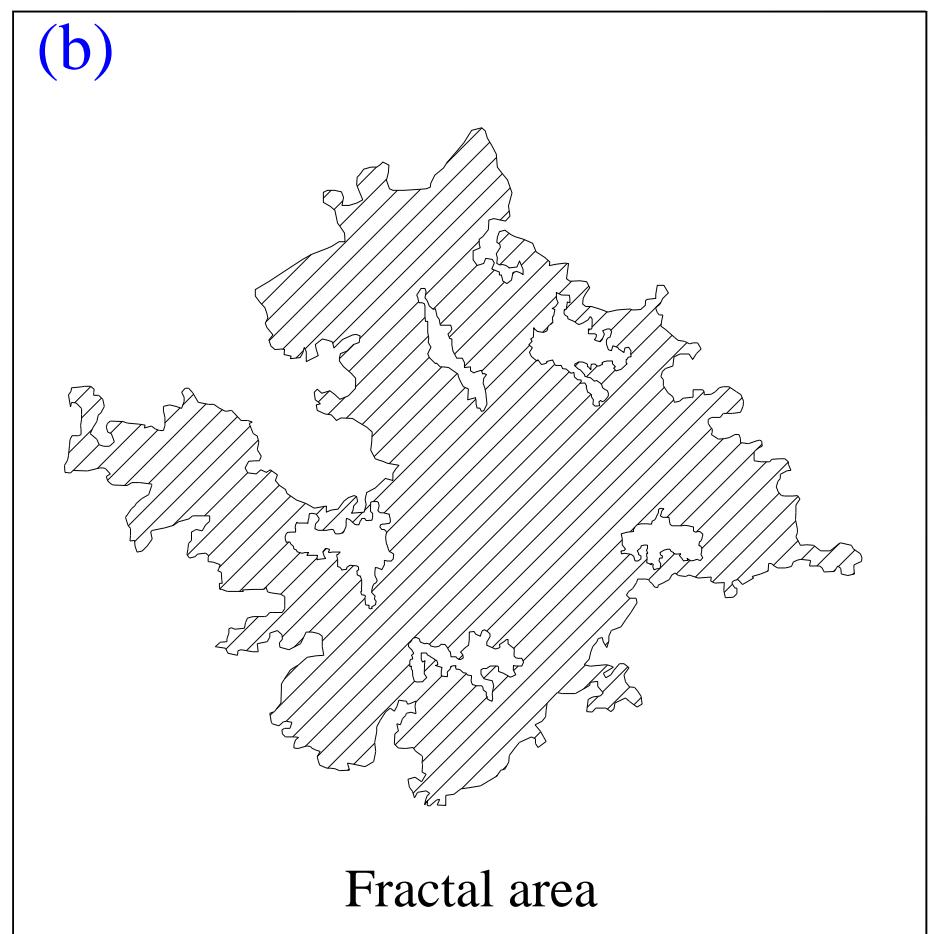

a. Euclidean urban area (Batty and Longley, 1994)  b. Fractal urban area(Longley *et al*, 1991)

**Figure 3 A sketch maps of the difference between Euclidean area and fractal area of cities**

**Note**: Figure 3(a) illustrates a Euclidean area, but figure 3(b) does not show a real fractal area. A real fractal has no area, and a prefractal region should include a hierarchy of urban patches. This is just a diagrammatic sketch. Taking into account the holes in the fractal area, Wentz (2000) defined a perforation index for shape description.

Based on the fractal models shown above, the lurking scaling in the shape indexes can be brought to light. Two shape indexes depend on the scaling of fractal form and boundary. Substituting equations (10) and (11) into the formulae of circularity ratios (CiR) yields

$$CiR_1 = \frac{4A}{P^2} = \frac{4A_1\varepsilon^{2-D_b}}{(P_1\varepsilon^{1-D_f})^2} = \frac{4A_1}{P_1^2}\varepsilon^{2D_f-D_b}, \tag{15}$$

$$CiR_2 = \frac{2\sqrt{\pi A}}{P} = \frac{2\sqrt{\pi A_1\varepsilon^{2-D_f}}}{P_1\varepsilon^{1-D}} = \frac{2\sqrt{\pi A_1}}{P_1}\varepsilon^{D_b-D_f/2}. \tag{16}$$

The two equations suggest that the circularity ratios depend on the linear size of measurement, and both the boundary dimension and form dimension influence the index values.

The majority of shape indexes depend on the scaling of fractal form. Substituting equation (10) into the formulae of compactness ratios (CoR), form ratio (FR), ellipticity index (EI), elongation ratio (ER), and the second radial shape index (RSI) yields

$$CoR = \frac{A}{A'} = \frac{4A}{\pi L^2} = \frac{4A_1}{\pi L^2}\varepsilon^{2-D_f} , \tag{17}$$

$$FR = \frac{A}{L^2} = \frac{A_1}{L^2}\varepsilon^{2-D_f} , \tag{18}$$

$$EI = \frac{L}{2A/[\pi(L/2)]} = \frac{\pi L^2}{4A} = \frac{\pi L^2}{4A_1}\varepsilon^{D_f-2} , \tag{19}$$

$$RSI_2 = \frac{A}{\sqrt{2\pi\int d_i^2 \mathrm{d}x\mathrm{d}y}} = \frac{A_1}{\sqrt{2\pi\int d_i^2 \mathrm{d}x\mathrm{d}y}}\varepsilon^{2-D_f} . \tag{20}$$

These equations imply that the above four shape indexes depend on the linear size of measurement, and only the form dimension influences the index values.

The similarities and differences between these spatial measurements are demonstrated. According to the scaling processes, shape indexes can be divided into three categories (Table 3). The first category includes two types of circularity ratios (CiR), which depend on the scale of measurement. When the linear size of ruler becomes very small, the perimeter approach infinity, and the values of indexes approach zero. The scaling in both perimeter and area of a city is involved in these shape indexes. The second category includes two types of compactness ratios (CoR), form ratio (FR), ellipticity index (EI), and elongation ratio (ER), which are asymptotically close to constants along with scale change. As shown above, based on Feret's diameter, all these indexes can be mathematically transformed into one another. This suggests that these shape indexes are equivalent to each other in theory. Only the scaling in urban area is involved in these shape indexes. The third category includes elongation ratio (ER) and a kind of radial shape index (RSI). Based on fixed urban boundary, these types of shape indexes do not change over the linear size of measurement scale. They can be used as spatial measurements with characteristic scale in many cases. However, based on variable urban boundaries, these shape indexes are not determinate. In other word, scaling hides behind the identifying process of urban boundaries (Chen *et al*, 2015). In fact, even if the urban boundaries are fixed, the third type of shape indexes is not really certain or comparable owing to the various influence factors (Griffith *et al*, 1986).

**Table 3 The classification result of the classical shape indexes of geographic area**

| Boundary type | Scaling type | Shape index (SI) |
|---|---|---|
| **Fixed boundary** | Scaling in perimeter and area | Circularity ratio (I), CiR |
| | | Circularity ratio (II), CiR |
| | Scaling in area | Compactness ratio (I), CoR |
| | | Compactness ratio (II), CoR |
| | | Form ratio, FR |
| | | Ellipticity index, EI |
| | | Radial shape index (II), RSI |
| **Variable boundary** | Characteristic scale and latent scaling | Elongation ratio, ER |
| | | Radial shape index (I), RSI |

# 3. Fractal analyses

## 3.1 The shape indexes of Koch island

Koch's island can be used to illustrate the three levels of scaling in shape indexes. As indicated above, the Koch island is a metaphor of geographical area. A random Koch island bears an analogy with urban envelope. For simplicity, let's examine the regular Koch's island (Figure 1). Suppose that the area of initial triangle is one unit, that is, $A_0$=1. The ultimate area of Koch's island, as shown above, is 8/5. The side length of the initial triangle is

$$S=\left(\frac{2}{\sin(\frac{\pi}{3})}\right)^{1/2}, \tag{21}$$

where $S$ denotes side length. The Feret diameter is

$$F=\frac{S}{\cos(\frac{\pi}{6})}=\left(\frac{2}{\sin(\frac{\pi}{3})}\right)^{1/2}/\cos(\frac{\pi}{6}), \tag{22}$$

where $F$ refers to Feret's diameter. This suggests that $L$≈1.755. Due to rotational symmetry of Koch's island, the second axis is also $L'$≈1.755, and the distance of the $i$th vertex from centroid of the figure is $d_i=L/2$. The perimeter can be given by equation (2), in which $N_1$=1, $\varepsilon=S$, $S/3$, $S/9$,…, $S/3^i$ ($i$=0, 1, 2, …). It is easy to calculate the various shape indexes of Koch's island based on the above-mentioned concepts and numbers (Table 4).

Given a Koch island, the boundary is fixed, but the length is uncertain. Apparently, the perimeter of Koch's island is a fractal line, but the area is Euclidean area. Only one scaling process is involved with the computation of shape indexes. In this context, two shape indexes based on perimeter and area, i.e., the two circularity ratios (CiR1 and CiR2), depend on the linear size of spatial measurement. This is the type of uncertainty. Five shape indexes based on area and major axis (Feret's diameter), including two compactness ratios (CoR1 and CoR2), form ratio (FR), ellipicity index (EI), and the second type of radial shape index (RSI2), fast converge to fixed values. This is the type of convergence. Two shape indexes based on major axis, minor axis, or distance, including elongation ratio (ER) and the first type of radial shape index (RSI1), do not change over spatial scale. This is the type of constancy. The second and third types are reliable in the case of fixed fractal boundary and Euclidean area.

**Table 4 The shape index values of Koch island based on different spatial scales of measurement**

| **Scale** | **Basic measure** | | | | **Shape index** | | | | | | | |
|---|---|---|---|---|---|---|---|---|---|---|---|---|
| $\varepsilon$ | $A$ | $P$ | $L$ | $A'$ | CiR1 | CiR2 | CoR1 | CoR2 | EI | FR | ER | RSI |
| $S/3^{0}$ | 1.000 | 4.559 | 1.755 | 2.418 | 0.192 | 0.778 | 0.413 | 0.413 | 2.418 | 0.325 | 1 | 0 |
| $S/3^{1}$ | 1.333 | 6.079 | 1.755 | 2.418 | 0.144 | 0.673 | 0.551 | 0.551 | 1.814 | 0.433 | 1 | 0 |
| $S/3^{2}$ | 1.481 | 8.105 | 1.755 | 2.418 | 0.090 | 0.532 | 0.613 | 0.613 | 1.632 | 0.481 | 1 | 0 |
| $S/3^{3}$ | 1.547 | 10.807 | 1.755 | 2.418 | 0.053 | 0.408 | 0.640 | 0.640 | 1.563 | 0.503 | 1 | 0 |
| $S/3^{4}$ | 1.577 | 14.409 | 1.755 | 2.418 | 0.030 | 0.309 | 0.652 | 0.652 | 1.534 | 0.512 | 1 | 0 |
| $S/3^{5}$ | 1.590 | 19.212 | 1.755 | 2.418 | 0.017 | 0.233 | 0.657 | 0.657 | 1.521 | 0.516 | 1 | 0 |
| $S/3^{6}$ | 1.595 | 25.616 | 1.755 | 2.418 | 0.010 | 0.175 | 0.660 | 0.660 | 1.516 | 0.518 | 1 | 0 |
| $S/3^{7}$ | 1.598 | 34.154 | 1.755 | 2.418 | 0.005 | 0.131 | 0.661 | 0.661 | 1.513 | 0.519 | 1 | 0 |
| $S/3^{8}$ | 1.599 | 45.539 | 1.755 | 2.418 | 0.003 | 0.098 | 0.661 | 0.661 | 1.512 | 0.519 | 1 | 0 |
| $S/3^{9}$ | 1.600 | 60.718 | 1.755 | 2.418 | 0.002 | 0.074 | 0.661 | 0.661 | 1.512 | 0.519 | 1 | 0 |
| $S/3^{10}$ | 1.600 | 80.958 | 1.755 | 2.418 | 0.001 | 0.055 | 0.662 | 0.662 | 1.512 | 0.520 | 1 | 0 |
| $S/3^{15}$ | 1.600 | 341.155 | 1.755 | 2.418 | 0.000 | 0.013 | 0.662 | 0.662 | 1.512 | 0.520 | 1 | 0 |
| $S/3^{20}$ | 1.600 | 1437.625 | 1.755 | 2.418 | 0.000 | 0.003 | 0.662 | 0.662 | 1.511 | 0.520 | 1 | 0 |
| $S/3^{25}$ | 1.600 | 6058.141 | 1.755 | 2.418 | 0.000 | 0.001 | 0.662 | 0.662 | 1.511 | 0.520 | 1 | 0 |

## 3.2 The shape indexes of a growing fractal

Growing fractals can be employed to further clarify the three sorts of shape indexes. Box fractal is a well-known simple growing fractal, which was proposed by Jullien and Botet (1987) and popularized by Vicsek (1989). Thus it is also termed Vicsek's figure. It was often employed to illustrate city fractals and urban growth (Batty and Longley, 1994; Frankhauser, 1998; Longley *et* *al*, 1991; White and Engelen, 1993). Both the boundary and area are fractals, and the two fractal dimension values are $D_f$=$D_b$=ln(5)/ln(3)=1.465 (Figure 4). Suppose that the area of initial square is one unit, that is, $A_0$=1, then the side length of the square is 1, and the perimeter is $P_1$ =4. The fractal area can be described with equation (10), and the fractal perimeter can be described with equation (11). This suggests that the area goes down and down to infinitesimal, and the perimeter goes up and up to infinity. Using the formulae of shape indexes, we can compute varied index values based on different linear size of fractal copies (Table 5).

The boundary of this growing fractal is also fixed, but the length and area are both uncertain. As indicated above, the perimeter is a fractal line, and the area is a fractal set. Two scaling processes are involved with the calculations of shape indexes. Two shape indexes based on perimeter and area, that is, the two circularity ratios (CiR1 and CiR2), depend on the linear scales of measurement; Five shape indexes based on area and major axis (Feret's diameter), including two compactness ratios (CoR1 and CoR2), form ratio (FR), ellipicity index (EI), and the second type of radial shape index (RSI2), also depend on the spatial scales of measurement. Two shape indexes based on major axis, minor axis, or distance such as elongation ratio (ER) and the first type of radial shape index (RSI1) do not change over the linear size of spatial measurement. In this instance, only one type of the shape indexes is reliable for shape description.

**Table 5 The shape index values of a growing fractal based on different spatial scales of measurement**

| **Scale** | **Basic measure** | | | | **Shape index** | | | | | | | |
|---|---|---|---|---|---|---|---|---|---|---|---|---|
| $\varepsilon$ | $A$ | $P$ | $L$ | $A'$ | CiR1 | CiR2 | CoR1 | CoR2 | EI | FR | ER | RSI |
| **$1/3^0$** | 1.000 | 4.000 | 1.414 | 1.571 | 0.250 | 0.886 | 0.637 | 0.637 | 1.571 | 0.500 | 1 | 0 |
| **$1/3^1$** | 0.556 | 6.667 | 1.414 | 1.571 | 0.050 | 0.396 | 0.354 | 0.354 | 2.827 | 0.278 | 1 | 0 |
| **$1/3^2$** | 0.309 | 11.111 | 1.414 | 1.571 | 0.010 | 0.177 | 0.196 | 0.196 | 5.089 | 0.154 | 1 | 0 |
| **$1/3^3$** | 0.171 | 18.519 | 1.414 | 1.571 | 0.002 | 0.079 | 0.109 | 0.109 | 9.161 | 0.086 | 1 | 0 |
| **$1/3^4$** | 0.095 | 30.864 | 1.414 | 1.571 | 0.000 | 0.035 | 0.061 | 0.061 | 16.490 | 0.048 | 1 | 0 |
| **$1/3^5$** | 0.053 | 51.440 | 1.414 | 1.571 | 0.000 | 0.016 | 0.034 | 0.034 | 29.681 | 0.026 | 1 | 0 |
| **$1/3^6$** | 0.029 | 85.734 | 1.414 | 1.571 | 0.000 | 0.007 | 0.019 | 0.019 | 53.426 | 0.015 | 1 | 0 |
| **$1/3^7$** | 0.016 | 142.890 | 1.414 | 1.571 | 0.000 | 0.003 | 0.010 | 0.010 | 96.167 | 0.008 | 1 | 0 |
| **$1/3^8$** | 0.009 | 238.150 | 1.414 | 1.571 | 0.000 | 0.001 | 0.006 | 0.006 | 173.101 | 0.005 | 1 | 0 |
| **$1/3^9$** | 0.005 | 396.916 | 1.414 | 1.571 | 0.000 | 0.001 | 0.003 | 0.003 | 311.582 | 0.003 | 1 | 0 |
| **$1/3^{10}$** | 0.003 | 661.527 | 1.414 | 1.571 | 0.000 | 0.000 | 0.002 | 0.002 | 560.848 | 0.001 | 1 | 0 |
| **$1/3^{15}$** | 0.000 | 8507.290 | 1.414 | 1.571 | 0.000 | 0.000 | 0.000 | 0.000 | 10597.598 | 0.000 | 1 | 0 |
| **$1/3^{20}$** | 0.000 | 109404.449 | 1.414 | 1.571 | 0.000 | 0.000 | 0.000 | 0.000 | 200248.826 | 0.000 | 1 | 0 |
| **$1/3^{25}$** | 0.000 | 1406950.220 | 1.414 | 1.571 | 0.000 | 0.000 | 0.000 | 0.000 | 3783837.741 | 0.000 | 1 | 0 |

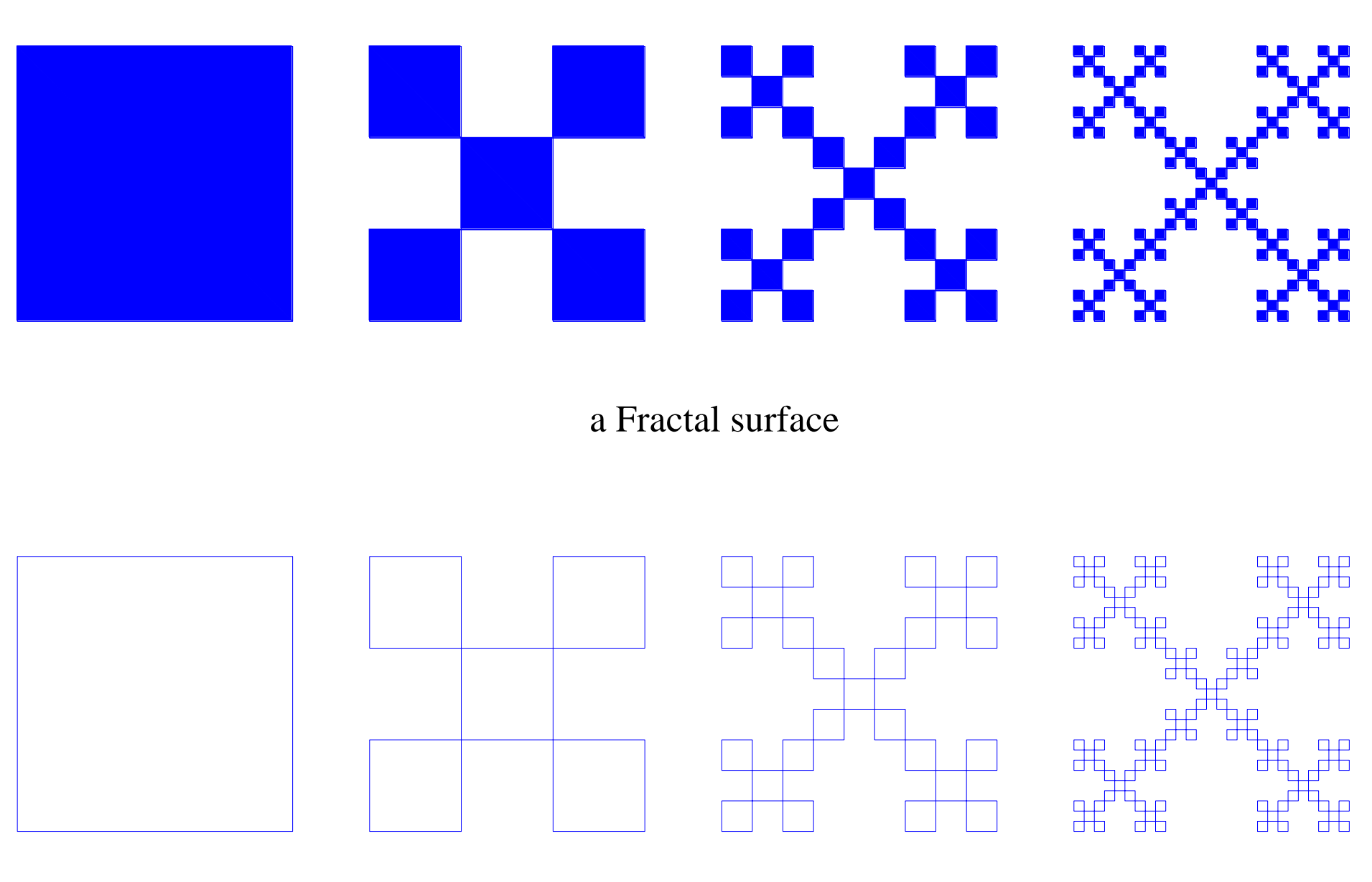

a Fractal surface

b Fractal boundary

**Figure 5 A growing fractal and its self-similar boundary line (The first four steps, by** Vicsek, 1989, Longley *et al*, 1991, and Frankhauser, 1998**)**

## 4. Questions and discussion

The case analyses based on two types of regular fractals indicate three types of shape indexes. If a perimeter such as urban boundary can be exactly delineated, some shape indexes are influenced by the scaling of fractal form and boundary. For example, the elongation ratio (ER) and the first type of radial shape index (RSI) are based on major axis and minor axis, and free of scaling. Therefore, it is more advisable to use this kind of indexes to describe the characteristic urban shape. For the box fractal, both the area and boundary depend on measurement scale and are not determinate; for Koch's island, the area is definite, but the perimeter is not determinate. However, for both the box fractal and Koch's island, the major axis and minor axis are clear. As a result, the shape indexes independent of area and perimeter are free of scaling given certain urban envelope, e.g., ER and RSI1; the shape indexes independent of perimeter are free of scaling given certain urban envelope and urban area, e.g., CoRs, FR, EI, and RSI1; the shape indexes based on perimeter and area are involved in scaling, e.g., CiRs. Real cities are more complex than any fractal models. It is very difficult to find a clear boundary for a real city. Spatial search method can be employed to delineate urban boundary curves. However, changing search radius yields different boundary curves (Chen *et al*, 2015). This suggests that the major axis and minor axis are not actually determinate in many cases. If so, we should find new ways of defining urban shape indexes. Despite this, partial urban shape indexes such as elongation ratio (ER) and the first type of radial shape index (RSI) are available for depicting urban figuration.

Scientific research is always involved in three worlds: real world, mathematical world, and computational world. Real fractals, regular ones or random ones, can only be found in the mathematical world, while all what is called fractals in the real world are in fact random prefractals (Addison, 1997; Ghanbarian-Alavijeh *et al*, 2011). For the prefractals, fractal properties only take on within certain scaling range (Chen, 2013). Despite this, we can study the prefractals such as cities using the concepts of fractal geometry. This involves the so-called computational world (Casti, 1996). Both the real world and mathematical world are objective, but the computational world is not so objective. There are many subjective algorithms, concepts, measurement, methods, and so on, in the computational world. Where the methods are concerned, there is no best, only better can be found. Consequently, for the same phenomenon, we have different approaches to describing and analyzing it. In practice, if shape indexes such as circularity ratios cannot be effectively estimated, we can adopt boundary dimension to characterize urban shapes. Treating urban shape as a random prefractal fractal, the perimeter is actually finite. It was demonstrated that, for the given perimeter ($P$), the circularity ratios ($CiR$s) of urban shape is the exponential function of the reciprocal of the boundary dimension ($D_b$) (Chen, 2011). The relation between the boundary dimension and Miller's circularity ratio can expressed as

$$CiR1=\frac{1}{k\pi P^2}\exp(\frac{2}{D_b}\ln P)\,, \tag{23}$$

where $2\ln P$ and $1/(k\pi P^2)$ are two parameters, $\pi$=3.1416, and $k$ is a constant. Accordingly, the relation between the Richardson's circularity ratio and boundary dimension is as below

$$CiR2=\frac{1}{\sqrt{k}P}\exp(\frac{1}{D_b}\ln P)\,, \tag{24}$$

where $\ln P$ and $1/(k^{1/2}P)$ are two parameters. It is easy to testify the two parameter relations. Using the fractal measure relation, equation (13), and the formula of Richardson's circularity ratio shown in Table 1, Song *et al* (2012) estimated the boundary dimension values ($D_b$) and second type of circularity ratio (CIR2) of 33 mining cities of China (Table 6). It is easy to compute the first type of circularity ratio by the formula as follows, $CiR1=CiR2/\pi$ (Chen, 2011). Applying equations (23) and (24) to the datasets of Chinese mining cities by means of least squares calculations yields two models as follows

$$CiR1=\frac{1}{1070104.4357}\exp(\frac{14.7508}{D_b}),$$

$$CiR2=\frac{1}{583.6307}\exp(\frac{7.3754}{D_b}).$$

The goodness of fit is about $R^2=0.9192$, the average perimeter is about $P=1596.2521$, the perimeter is $k=0.1337\approx 1/\ln P$. In the process of parameter estimation, one of mining city, Fushun, is eliminated as an outlier because of its standardized residual $e=-3.7930<-2$. In statistical analysis, if the absolute value of a standardized residual is greater than 2, the corresponding observational number can be treated as an exceptional value by the confidence level of 95%. These modeling results lend further support to the suggestion that there is a mathematical and numerical relation between shape indexes and fractal dimension of urban form.

**Table 6 The boundary dimension and circularity ratios values of China's 33 mining cities in 2006**

| City | Boundary dimension | | Circularity ratios | | Standardized residuals |
|---|---|---|---|---|---|
| | $D_1$ | $R^2$ | CiR1 | CiR2 | $e_i$ |
| **Anshan** | 1.441 | 0.978 | 0.0204 | 0.2530 | -0.7084 |
| **Baiyin** | 1.283 | 0.985 | 0.0667 | 0.4576 | -1.1089 |
| **Benxi** | 1.593 | 0.980 | 0.0072 | 0.1503 | -0.8084 |
| **Daqing** | 1.555 | 0.978 | 0.0158 | 0.2227 | 1.0350 |
| **Datong** | 1.413 | 0.983 | 0.0340 | 0.3269 | 0.3012 |
| **Dongying** | 1.443 | 0.982 | 0.0225 | 0.2658 | -0.3288 |
| **Fushun** | *1.526* | *0.975* | *0.0045* | *0.1186* | *-3.7930* |
| **Fuxin** | 1.448 | 0.983 | 0.0318 | 0.3159 | 0.9450 |
| **Hebi** | 1.498 | 0.981 | 0.0240 | 0.2747 | 1.1862 |
| **Hegang** | 1.510 | 0.981 | 0.0154 | 0.2199 | -0.0253 |
| **Huaibei** | 1.487 | 0.979 | 0.0201 | 0.2511 | 0.3365 |
| **Huainan** | 1.525 | 0.982 | 0.0155 | 0.2209 | 0.3344 |
| **Jixi** | 1.574 | 0.979 | 0.0085 | 0.1635 | -0.6312 |
| **Jinchang** | 1.251 | 0.989 | 0.1395 | 0.6619 | 0.3367 |
| **jincheng** | 1.309 | 0.986 | 0.0639 | 0.4480 | -0.4614 |
| **Kelamayi** | 1.284 | 0.989 | 0.1008 | 0.5627 | 0.3004 |
| **Liaoyuan** | 1.509 | 0.977 | 0.0229 | 0.2683 | 1.2773 |
| **Liupanshui** | 1.444 | 0.975 | 0.0193 | 0.2462 | -0.8163 |
| **Maanshan** | 1.337 | 0.987 | 0.0779 | 0.4946 | 1.0138 |
| **Panzhihua** | 1.546 | 0.986 | 0.0102 | 0.1793 | -0.6038 |
| **Panjin** | 1.509 | 0.981 | 0.0152 | 0.2183 | -0.0971 |
| **Pingdingshan** | 1.517 | 0.980 | 0.0204 | 0.2533 | 1.0723 |
| **Pingxiang** | 1.358 | 0.982 | 0.0468 | 0.3833 | -0.0972 |
| **Puyang** | 1.476 | 0.981 | 0.0251 | 0.2806 | 0.8228 |
| **Qitaihe** | 1.546 | 0.988 | 0.0136 | 0.2064 | 0.3360 |
| **Shizuishan** | 1.256 | 0.989 | 0.1482 | 0.6823 | 0.7010 |
| **Shuangyashan** | 1.597 | 0.984 | 0.0119 | 0.1935 | 0.9597 |
| **Shuozhou** | 1.326 | 0.981 | 0.0476 | 0.3867 | -0.9449 |
| **Songyuan** | 1.468 | 0.981 | 0.0304 | 0.3092 | 1.2804 |
| **Tangshan** | 1.476 | 0.982 | 0.0164 | 0.2271 | -0.5881 |
| **Tongchuan** | 1.426 | 0.982 | 0.0234 | 0.2711 | -0.6186 |
| **Wuhai** | 1.557 | 0.981 | 0.0112 | 0.1876 | -0.0682 |
| **Yangquan** | 1.489 | 0.983 | 0.0152 | 0.2187 | -0.5391 |

**Note**: The data come from Song *et al* (2012). Let $1/D_b$ be an independent variable, and ln($CiR$) be a dependent variable, we can make a linear regression using all the data points. Thus, the standardized residual value of Fushun is $e$=-3.7930. The standard deviation of the standardized residuals is 1. The normal standardized residuals come between -2 and 2. So, Fushun is removed as an outlier before estimating the final model parameters.

In literature, shape indexes are chiefly regarded as spatial measurements with characteristic scales. In many cases, shape indexes can be used to reflect the characters of size and shape of cities (Haggett *et al*, 1977; Lin, 1998; Song *et al*, 2012; Wang *et al*, 2005). However, the comparability of different shape indexes is influenced by scaling property of urban form and growth. Compared with previous studies on shape indexes, this paper focuses on the difference between scaling and characteristic scales behind the spatial measurements. One of the shortcomings in this study is that no empirical analysis is made for the classification results of shape indexes. In fact, a series of papers are composed for researching urban description. This work is a pure theoretical exploration, aiming at normalizing several correlated shape indexes by Feret's diameter and classifying varied shape indexes by scaling idea. It is difficult to clarify three types of empirical cases of urban shapes. A companion paper is devoted to making a positive study on the scaling in shape indexes of real cities.

## 5. Conclusions

Today, we know that the uncertainty of spatial measurement and shape index evaluation proceeds from the scaling property in urban form. There are various new and old shape indexes, and this paper is devoted to examining the classical ones. The main conclusions can be reached as follows. **Firstly, given a determinate urban boundary, the shape indexes can be divided into three types: uncertain type (two scaling processes), half-certain type (one scaling process), and certain type (characteristic scale).** The first type such as circularity ratios (CiRs) is based on the ratio of urban area to square of urban perimeter. Both the boundary and form can be treated as fractals, and urban perimeter and area depend on the scale of spatial measurement. At least, the perimeter depends on the measurement scale because urban boundary bears an analogy with a fractal line. Therefore, this type of shape indexes cannot be exactly calculated. The second type such as compactness ratios (CoR) and form ratio (FR) is based on the ratio of urban area to square of major axis length. If the urban area is treated as a Euclidean surface, this type of indexes can be effectively computed. If we consider the internal structure of urban land use and irregular patterns within the boundary, urban area will depend on the scale of measurement. In this case, this type of shape indexes cannot be precisely worked out. The third type such as Elongation ratio (ER) is based on the ratio of the longest (major) axis length to the secondary (minor) axis length, or based on the distances of vertexes from centroid of a figure. Since the urban boundary is regarded as clearly defined, the lengths of main axes or distances are determinate in theory. Thus, this type of indexes is relatively effective. **Secondly, the scaling of shape indexes can be investigated from two aspects and three levels.** If the urban boundary is treated as clearly delineated, the scaling can be considered at two levels. One level is that only the boundaries is treated as fractal lines (one scaling process), the area within the boundary is regarded as Euclidean surface. If so, the shape indexes involving with perimeter will be uncertain. The second level is that both the boundary and the area within the boundary are regarded fractals (two scaling processes). If so, the shape indexes involving with perimeter or area will be uncertain. However, for the cities in the real world, urban boundaries cannot be clearly defined. This involves the second aspect and the third level of scaling in shape indexes. If urban boundaries depend on the scale of measurement and become variable, all the shape indexes will be of scale dependence, without exception. In practice, the urban boundary at a given time can be subjectively defined by certain statistic criteria, and thus the shape indexes independent of perimeter and area will be advisable for spatial description of urban shape. In the simplest case, urban boundary is clearly defined and urban area is treated as Euclidean surface, thus only the two circularity ratios depend on the scale of spatial measurement.

## Acknowledgements

This research was sponsored by the National Natural Science Foundation of China (Grant No. 41671167). The supports are gratefully acknowledged.